\documentclass[10pt,letterpaper]{article}

\usepackage{opex3}
\usepackage{siunitx}
\usepackage{amsmath}
\usepackage{xcolor}
\usepackage{cite}
\usepackage{hyperref}

\definecolor{dkgreen}{rgb}{0,0.6,0}
\definecolor{gray}{rgb}{0.5,0.5,0.5}
\definecolor{mauve}{rgb}{0.58,0,0.82}

\begin{document}

\title{A conjugate gradient minimisation approach to generating holographic traps for ultracold atoms}
\author{Tiffany Harte,$^{1,2}$ Graham D. Bruce,$^{1,3}$ Jonathan Keeling$^1$ and Donatella Cassettari$^{1,*}$}
\address{$^1$School of Physics and Astronomy, University of St Andrews,
St Andrews, Fife KY16 9SS, UK \\ $^2$Clarendon Laboratory, University of Oxford, Parks Road, Oxford, OX1 3PU, UK \\ $^3$SUPA Dept. of Physics, University of Strathclyde, 107 Rottenrow, Glasgow, G4 0NG, UK}
\email{$^*$dc43@st-andrews.ac.uk}

\begin{abstract}
	Direct minimisation of a cost function can in principle provide a versatile and highly controllable route to computational hologram generation.  However, to date iterative Fourier transform algorithms have been predominantly used.  Here we show that the careful design of cost functions, combined with numerically efficient conjugate gradient minimisation, establishes a practical method for the generation of holograms for a wide range of target light distributions. This results in a guided optimisation process, with a crucial advantage illustrated by the ability to circumvent optical vortex formation during hologram calculation. We demonstrate the implementation of the conjugate gradient method for both discrete and continuous intensity distributions and discuss its applicability to optical trapping of ultracold atoms. \end{abstract}


\bibliographystyle{osajnl}
\bibliography{StAndrewsConjugateGradient}

\section{Introduction}

In recent years there has been extraordinary progress in cold atom physics and its applications in fields such as quantum computation and simulation of condensed-matter systems, precision measurements, and matter-wave interferometry~\cite{SimRev, IntRev}. 
In this context, arbitrary time-dependent optical trapping potentials are particularly appealing, with a variety of geometries including toroids and ring lattices already realised by acousto-optic or holographic means~\cite{henderson_experimental_2009, Kuhn2012}. Experiments have been performed with discrete arrays of optical dipole traps, loaded with either thermal atoms~\cite{Bergamini, Kuhn2012} or quantum degenerate atomic gases~\cite{boyer_dynamic_2006, henderson_experimental_2009, Esslinger}, in which individual trapping sites can be moved, addressed and manipulated. Important too are continuous trapping geometries: the primary subject of the present work are extended (as opposed to diffraction-limited) power-law potentials, proposed both as a static supplement to a trapping potential to cancel unwanted external potentials~\cite{AOD_2013}, and in a dynamic sequence as a tool for the efficient production of Bose-Einstein condensates~\cite{Bruce_powerlaw_2011}. Other interesting continuous potentials include engineered waveguides with dynamic bright regions, shown to be suitable for studies of BEC superfluidity~\cite{Bruce_ring_2011}.

Technologies employed so far in the realisation of these arbitrary optical trapping patterns include acousto-optic deflection of a laser beam to produce either a composite static intensity distribution~\cite{AOD_2013} or a rapidly-scanned profile~\cite{henderson_experimental_2009, Arnold_BlueDetunedScanning, Esslinger}, digital micro-mirror devices (DMDs)~\cite{Kuhn2012}, and computer-generated holograms implemented with phase-only spatial light modulators (SLMs)~\cite{Dholakia, Bergamini, boyer_dynamic_2006, pasienski, Bruce_ring_2011, Bruce_powerlaw_2011, Gaunt_2012, Lee_14}.  The high phase-resolution available in phase-only SLMs offers significant advantages for versatility of the accessible trapping patterns, though at the cost of lower switching speed between frames if compared to acousto-optic modulators and digital mirror devices.  However, with new technologies currently being developed for grey-scale phase-only SLMs with kHz refresh rates~\cite{Warde}, this versatility may become accessible at sufficiently high update rates for high-speed dynamic manipulation of trapped atoms.  The primary challenge of phase-only SLMs is the computational complexity inherent in reproducing the target intensity distribution on the trapping plane. This paper demonstrates an alternative reliable and efficient method to address this problem.

Our investigation concerns an SLM consisting of $256\times256$ programmable pixels; each pixel is able to impose a phase retardation between $0$ and $2\pi$ in steps of $2\pi /256$ on an incident laser beam. 
The resulting digital hologram is calculated to reconstruct a given target intensity pattern in the far field, or equivalently in the focal plane of a lens, in which atoms will be trapped.  The calculated phase mask $\phi_{pq}$ and the incident laser field $A_{0}S_{pq}$, with indices \textit{p} and \textit{q} denoting pixel position, determine the SLM--plane electric field:
\begin{equation}
\label{eqn:Ein}
E_{in} = A_{0} S_{pq}\exp\left(i \phi_{pq}\right).
\end{equation}                                                   

We express this electric field as an array of \textit{N} pixels; propagation through focussing optics can be calculated by a fast Fourier transform.  The electric field in the output plane is therefore given by
\begin{equation}
\label{eqn:Eout}
E_{out} = \frac{A_{0}}{N} \sum_{pq} S_{pq} \exp\left(i \phi_{pq}\right) \exp\left(- \frac{2 \pi i}{N}\left(pn+qm\right)\right),
\end{equation}

 with output--plane coordinates denoted by \textit{n} and \textit{m}.  As only the modulus of $ E_{out}$ is relevant for optical trapping, we have output--plane phase freedom: consequently the phase $\phi_{pq}$ required to recreate a given target intensity is not unique, and solutions are found numerically. The general problem of phase retrieval, which includes both the above case of laser beam shaping where the modulus of the field is known in both planes, and the related problem of image reconstruction where the field modulus is known only in the output plane, can be solved using a variety of methods. These fall broadly into two categories: iterative Fourier transform algorithms (IFTAs), and algorithms based on the minimisation of a cost function.  The IFTA calculation encourages convergence from an initial phase guess to one yielding the target intensity using successive Fourier transforms between SLM and output planes, imposing the known or desired electric field amplitude at each step. For the purpose of atom trapping in arbitrary geometries, where smoothness is a primary consideration (as corrugations of the trapping potential cause fragmentation of cold atom clouds \cite{corrugation}), the best results so far have been achieved by the mixed-region amplitude freedom (MRAF) variant of the IFTA~\cite{pasienski,Bruce_ring_2011, Bruce_powerlaw_2011, Gaunt_2012}. In MRAF, the output plane is divided into two regions: a signal region in which the intensity is restricted to match the target intensity pattern, and a noise region in which the intensity is unconstrained. This separation allows for increased accuracy and smoothness in the signal region, leading to computed intensity patterns with residual root-mean-square (RMS) errors of less than a few percent.  
A secondary consideration in designing optical traps is the light-usage efficiency of the computer-generated hologram.  One motivation for the use of phase-only spatial light modulators rather than amplitude modulators is that the former do not deliberately remove light from the incident beam.  However, the MRAF algorithm gains accuracy by deliberately lowering this efficiency.
Furthermore, lacking a minimisation principle, IFTA approaches provide no guarantee of converging, and their final state can be highly dependent on the initial phase pattern used.

In contrast, cost function minimisation algorithms are inherently more directional than IFTAs: the cost function encodes all constraints and desired properties of the intensity pattern, and can be designed with terms accounting for specific output plane features (such as high light-usage efficiency) in addition to adherence to the target intensity profile.  
Within this category, established beam shaping methods include genetic algorithms and direct search algorithms, both of which are less computationally efficient than IFTAs \cite{pasienski}. Genetic algorithms~\cite{genetic_Mitchell} seek the global minimum of the cost function, and as such are computationally demanding but accurate. Direct search algorithms~\cite{directsearch, boyer_dbs}, in which SLM pixel values are sequentially altered with only changes reducing the cost function being retained, are limited to just a few phase levels due to computational intensity, and as such work well for simple targets but struggle to reproduce more intricate patterns.

In this paper, we consider an alternative approach to the beam shaping problem, in which the cost function is minimised by a conjugate gradient local search algorithm. 
Conjugate gradient minimisation, a well--established method for minimising high-dimensional smooth functions, is widely used in contexts such as electronic structure \cite{payne}.
Here we find that this approach successfully combines computational efficiency and algorithm versatility, allowing the accurate reproduction of a variety of target intensity profiles relevant for optical trapping of atoms.  The simplest cost function we study, a least-squares difference from the desired pattern, leaves localised defects which have low cost but present significant problems for atom trapping.  However, these defects can be removed by systematically modifying the cost function, and we discuss the forms of cost functions required to eliminate them.  This flexibility in cost function definition, a useful feature common to all minimisation algorithms, also allows us to go beyond the simple definitions of signal and noise regions, and to fine-tune our algorithm for different experimental requirements simply by adding cost function terms and applying different weightings across the output plane. 

Laser beam shaping via an algorithm that relies entirely on gradient-based local search has been relatively unexplored so far. Examples of this are found in~\cite{cg_98, cg_98_2, colour}, where conjugate gradient minimisation is used to generate pseudo-non diffractive beams in which the target is a given axial intensity distribution. Gradient-based local search techniques are also used as part of more complex algorithms, such as hybrid algorithms for beam shaping in which they are combined with genetic algorithms to increase the reliability in locating the global minimum for the generation of flat top laser beams~\cite{dpe, beam_shaping_hybrid}. The present work shows that gradient-based local search can be applied to a much wider range of beam shaping problems, and that it can be applied on its own: finding a local minimum is sufficient for a good reproduction of a variety of targets.

\section{The conjugate gradient calculation method}

\begin{figure}[!htb]
	\centering
	\includegraphics[scale=.6]{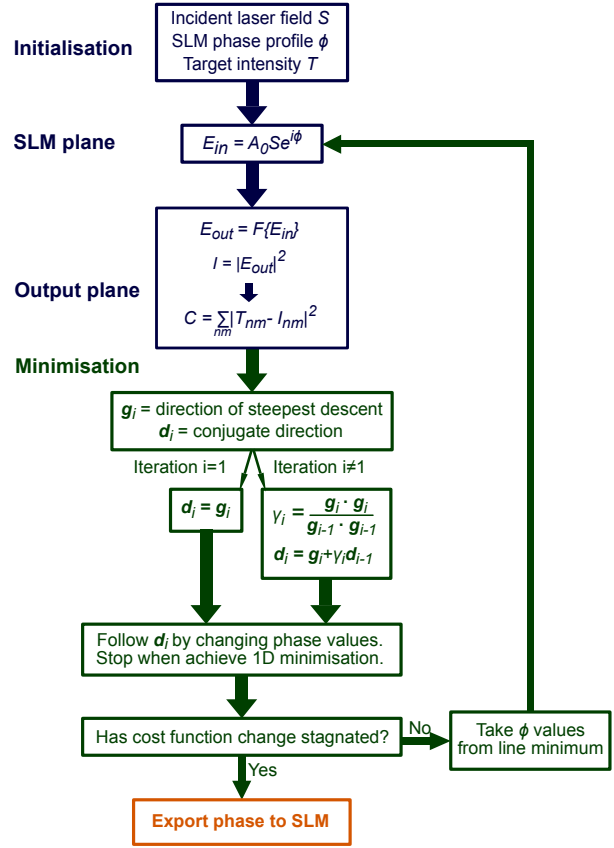}
	\caption{Block diagram illustrating the conjugate gradient minimisation approach to hologram solution. }
	\label{fig:cgflow1}
	\end{figure}

At the heart of this method is quantifying the error between target and predicted intensity by defining a cost function $C$.  Minimisation of $C$ is performed over all SLM pixel phase values, while computational efficiency is ensured by incorporating gradient information. Figure~\ref{fig:cgflow1} illustrates the application of conjugate gradient minimisation to hologram calculation.

The process is initialised by defining a target intensity distribution, an incident laser field amplitude, and an initial phase guess.  This may be the summation of analytically-expressed phase patterns to form an educated guess, e.g. quadratic and linear phase gradients giving expansion and position-offset of the intensity in the output plane~\cite{pasienski}, or it may be a random array taking values between $0$ and $2\pi$, with each element corresponding to an SLM pixel. 

Weighting arrays are also defined during this initialisation stage.  This weighting process is significantly more flexible than the simple definition of signal and noise regions characteristic of the MRAF IFTA.   Output plane regions can be arbitrarily weighted according to their importance: we can allocate a pixel-dependent prefactor to individual cost function terms according to their relative importance in different output plane regions. For the remainder of this paper, the high-intensity output region forming the trapping pattern is referred to as the trapping region, with the signal and noise regions retaining the same meanings as in MRAF: the signal region is the trapping region plus some border which will remain devoid of light, while the noise region is the remainder of the plane where light may be deposited without adversely affecting the trapping potential. For example, the cost function can be allocated a larger weighting, so greater importance, in the trapping region than the remainder of the output plane, while the highest intensity parts of this region preferentially seen by the atoms can be given yet more prominence than those of lower intensity. With our cost function approach, we can provide a smooth transition between signal and noise region weightings which reduces noise accumulation at the signal region border. By comparison, a typical light pattern calculated using MRAF places much of the noise region light at the boundary of the signal and noise regions.
	
	Upon each iteration, the output electric field corresponding to the current phase profile is calculated, with phase array components generated by the minimisation routine. This multidimensional minimisation is composed of one-dimensional steps, each seeking to minimise the cost function by changing phase values. The initial step minimises the cost function in the local gradient direction; subsequent consecutive minimisation directions are conjugate and independent, to avoid repetition of minimisation directions~\cite{numericalrecipes, agonisingpain}.  Conjugate directions $\mathbf{d}$ are those satisfying \cite{payne, agonisingpain}:
	\begin{equation}
		\mathbf{d}_{i} \cdot H_{C} \cdot \mathbf{d}_{j} = 0,
	\end{equation}
	where $H_{C}$ is the Hessian matrix of the cost function.  However, the power of the conjugate gradient descent approach is that the Hessian does not need to be calculated explicitly, but is rather built up by application of the following procedure.  The $i$th direction is calculated using \cite{payne}:
	\begin{equation}
		\mathbf{d}_i = \mathbf{g}_i+\gamma_{i}\mathbf{d}_{i-1}
	\end{equation}
	where $\mathbf{g}_{i}$ is the direction of steepest descent at the termination point of step $i-1$ and $\gamma_{i}$ is the scalar
	\begin{equation}
		\gamma_{i}=\frac{\mathbf{g}_{i} \cdot \mathbf{g}_{i}}{\mathbf{g}_{i-1} \cdot \mathbf{g}_{i-1}}.
	\end{equation}
	
	The fast Fourier transforms used in the calculations map a $N \times N$ array in the SLM plane onto a $N \times N$ array in the output plane.  If $N$ is chosen to be the number of pixels in the SLM, then the size of each output plane pixel is exactly the diffraction limit of the system.  Aliasing in the output plane is avoided by selecting a value for $N$, in accordance with the Nyquist criterion, of twice the number of pixels in the SLM \cite{Johansson}. Therefore the phase array is surrounded by zeroes to double its size and optimise output plane sampling at each iteration. Correspondingly enlarging the target array, the resolution of the cost calculation is optimised. Iteration continues until the difference between consecutive cost values stagnates: a minimum of the chosen cost function has been located.  Our calculations make use of the libatoms library~\cite{libatoms}.
		
\section{Versatility via cost function definition}

	The cost function should be such that its minimum corresponds to the desired pattern.  There are however a number of other features required for the algorithm to operate efficiently: it should be efficient to evaluate derivatives of the cost function; the function should not have local minima which give poor trapping profiles.  Indeed, as we will see below, a naive choice of cost function leads to local minima containing optical vortices.  Furthermore, the conjugate gradient approach assumes an approximately quadratic function~\cite{numericalrecipes}. 
	A simple cost function may purely concern target reproduction accuracy, expressed as a sum over output plane pixels $\left(n,m\right)$ of differences between target $T_{nm}$ and calculated output intensity:
	\begin{eqnarray}
		\label{eqn:cost1}
		C = \sum_{nm} \left(T_{nm} - \left|E_{out,nm}\right|^{2}\right)^{2} = \sum_{nm} \left(T_{nm} - \left|A_{0} \tilde{\psi}_{nm}\right|^{2}\right)^{2}
	\end{eqnarray}
		using
		\begin{equation}
			\tilde{\psi}_{nm} = \frac{1}{N} \sum_{pq} \psi_{pq} \exp\left(- \frac{2 \pi i}{N} (pn+qm)\right)
		\end{equation}
                with $\psi_{pq} = S_{pq} \exp\left(i \phi_{pq}\right)$, where indices \textit{p} and \textit{q} indicate general input plane pixels. $A_0$ is a free scale parameter allowing for the fact that the overall scale of the target potential is not necessarily matched to the laser power and $S_{pq}$ allows for the possibility of an input field with a slowly varying intensity, e.g. a Gaussian input beam. The cost function gradient required by the minimisation algorithm must be calculated with respect to $\phi_{rs}$, the phase value at a specific SLM pixel. For this cost function, the corresponding gradient is
		\begin{equation}
			\frac{\partial C}{\partial \phi_{rs}} =  4 A_{0}^{2} \textnormal{Re} \left(i \psi_{rs}^{*} X_{rs}\right)
		\end{equation}
		where
                \begin{equation}
                  X_{rs} = \frac{1}{N}\sum_{nm}\left[ \left(T_{nm} - \left|A_{0} \tilde{\psi}_{nm}\right|^{2}\right) \tilde{\psi}_{nm}^{*} \exp\left(\frac{2 \pi i}{N} \left(rn+sm\right)\right)\right].
                \end{equation}
		
	Additional terms in the cost function can incorporate experimentally relevant output plane features. Such features could, for instance, include noise suppression at the signal region boundary to aid trap loading, or in a dynamic sequence for real-time manipulation of trapped atoms, a cost function term could be introduced to reduce intensity fluctuations between consecutive frames in the sequence.  The application of additional cost function terms is illustrated here by the suppression of optical vortex formation.  We find that the cost function in Eq.~(\ref{eqn:cost1}) is effective for lattice distributions, but inadequate for large continuous patterns due to the emergence of optical vortices within the trapping region during calculation.  These vortices are characterised by a sudden drop in intensity coinciding with a local phase winding by a multiple of $2\pi$, and they arise because they can be initially beneficial to cost function reduction. However, their prevention is imperative to all hologram calculation schemes. From following the evolution of our conjugate gradient minimisation, it appears the local vorticity cannot change, and so these vortices can only be removed by annihilation of oppositely charged vortex pairs, or by moving vortices to regions of low intensity. 
	
\begin{figure}[!htb]
	\centering
	\includegraphics[scale=.35]{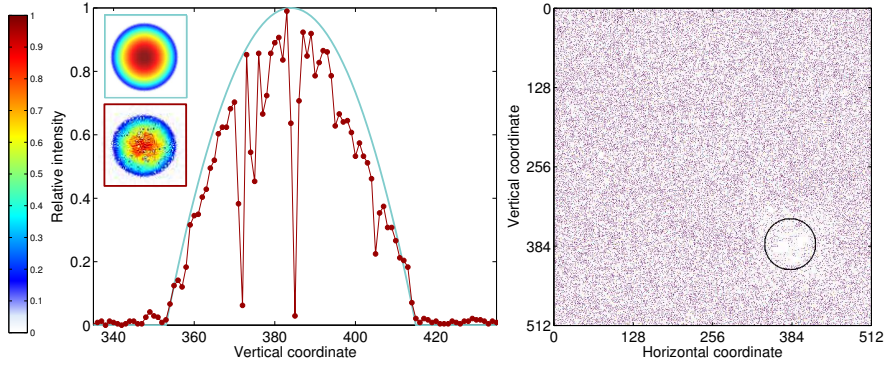}
	 	\caption{Second-order power-law trapping potential calculated using the cost function in Eq.~\ref{eqn:cost1}.  (left)~Two-dimensional profiles and vertical line profiles taken through the centre of the pattern for target (upper inset and cyan line) and calculated output (lower inset and red points).  The colorbar applies to the insets and to all subsequent figures.  The predicted output is distorted by deep optical vortices. These vortices inhibit further improvements in accuracy, resulting in a poor fit to the target intensity. The root-mean square (RMS) fractional error between target and predicted output is 26\%.  (right)~Vortex locations on the output plane, with red pixels indicating $2\pi$ phase windings and blue pixels $-2\pi$. The black circle indicates the trapping region. While the vortex density is reduced within this region in comparison to the remainder of the output plane, 232 are established within the trapping region with an insignificant fraction removed with further iterations due to the tangled phase contours.}
		\label{fig:vortices}
	\end{figure}

	Figure~\ref{fig:vortices} illustrates the vortices formed within the output plane of a second-order power-law intensity distribution calculated using the cost function defined in Eq.~(\ref{eqn:cost1}) with no regional weightings applied, starting from an educated guess.  Since the algorithm has identified a local minimum of the cost function, the observation of these vortices suggest that the cost function does not sufficiently penalise them.  Indeed, since the vortex cores are small, they only introduce a very localised deviation from the pattern.  Moreover, there is only a small change in cost function as vortices move through the pattern, and hence minimisation of this cost function does not effectively eliminate vortices once formed.  In minimising this cost function, vortex removal is principally achieved by gradually shifting them towards lower-intensity regions where their cost is reduced, but this is obstructed in regions of high vortex density where phase contours can become tangled~\cite{Senthilkumaran}. Vortex elimination is therefore only realistically achievable if their early formation is suppressed such that their numbers remain manageable.  Given that the cost function can be chosen at will (within the constraints given above), our approach to eliminating vortices becomes a question of choosing a \emph{better} cost function such that vortices do not remain frozen in the final pattern.
	
		A cost function that penalises large localised deviations more than the simple cost function in Eq. (\ref{eqn:cost1}) is
				\begin{equation}
			C_{t} =  \sum_{nm} \left(T_{nm} - \left|A_{0} \tilde{\psi}_{nm}\right|^{2}\right)^{t}. 
		\end{equation}

The higher the value of \textit{t}, the higher the cost of large discrepancies relative to small, increasing the cost contribution of trapping--region vortices. Fewer vortices persist, but at the expense of trap smoothness. In practice we find that powers higher than four produce too rough an intensity distribution with insufficient vortex improvements to justify this sacrifice.	The $C_{t}$ gradient is
		\begin{equation}
			\frac{\partial C_{t}}{\partial \phi_{rs}} =  2t A_{0}^{2} \textnormal{Re}\left(i \psi_{rs}^{*} X_{rs}\right).
		\end{equation}
		
	Alternatively, we can also specify cost functions that perform active smoothing by associating a cost with intensity variations between neighbouring pixels.  For example, to apply active smoothing over the four nearest-neighbour pixels, we use the cost function

		\begin{align}
			C_{s} &= \sum_{nm} \left[ \left(\left|\tilde{\psi}_{nm}\right|^{2}-\left|\tilde{\psi}_{n(m-1)}\right|^{2}\right)^{2}  + \left(\left|\tilde{\psi}_{nm}\right|^{2}-\left|\tilde{\psi}_{n(m+1)}\right|^{2}\right)^{2} \right.\nonumber\\ &\qquad \left. {} + \left(\left|\tilde{\psi}_{nm}\right|^{2}-\left|\tilde{\psi}_{(n-1)m}\right|^{2}\right)^{2}  + \left(\left|\tilde{\psi}_{nm}\right|^{2}-\left|\tilde{\psi}_{(n+1)m}\right|^{2}\right)^{2}\right].
		\end{align}

        The gradient is calculated in the same way for all four terms in $C_{s}$.  For instance, the gradient for the first term, $C_{s}^{(1)}~=~\sum_{nm} \left(|\tilde{\psi}_{nm}|^{2}-|\tilde{\psi}_{n(m-1)}|^{2}\right)^{2}$, is
		\begin{equation}
                        \frac{\partial C_{s}^{(1)}}{\partial \phi_{rs}} =  \frac{4}{N} \textnormal{Re}\left( i \psi_{rs} \left( X_{1rs}^{*} - X_{2rs}^{*}\right) \right)
		\end{equation}
		where 
		\begin{equation}
                  X_{1rs} = \frac{1}{N} \sum_{nm} \left[\left(\left|\tilde{\psi}_{nm}\right|^{2}-\left|\tilde{\psi}_{n(m-1)}\right|^{2}\right) \tilde{\psi}_{nm} \exp\left(\frac{2 \pi i}{N} (rn+sm)\right)\right]
		\end{equation}
		\begin{equation}
                  X_{2rs} = \frac{1}{N}\sum_{nm} \left[\left(\left|\tilde{\psi}_{nm}\right|^{2}-\left|\tilde{\psi}_{n(m-1)}\right|^{2}\right) \tilde{\psi}_{n(m-1)} \exp\left(\frac{2 \pi i}{N} (rn+sm)\right)\right].
		\end{equation}

\begin{figure}[!htb]
	\centering	
	{
		\includegraphics[scale=.35]{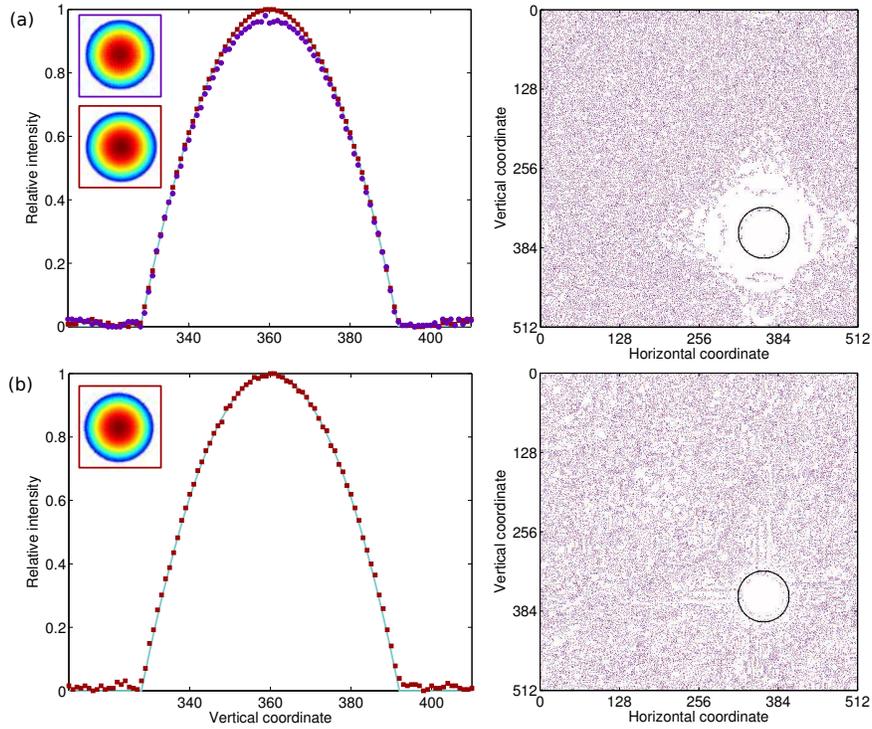}
		}
		\caption{Second-order power-law trapping potentials calculated with more sophisticated cost functions, starting with an educated guess phase.  (left)~Line profiles through pattern centre and two-dimensional profiles as insets. Calculated intensity patterns are denoted by points while the target profile is indicated with a cyan line.  (right)~output plane vortex map with the trapping region indicated by the black circle.  
		(a)~$C_{t=4}$ (purple, upper inset) and subsequent $C_{t=2}$ application (red, lower inset). Applying $C_{t=4}$ generates an approximate fit to the target while suppressing the initial vortex number. Subsequent smoothing using $C_{t=2}$ improves the accuracy and smoothness, and clears residual vortices from the trapping region. 
		(b)~A combination of $C_{t=2}$ and $C_{s}$ achieves both direct smoothing and high reproduction accuracy.}
			\label{fig:cterms}
	\end{figure}

	Sequential combination of $C_{t=4}$ and $C_{t=2}$ establishes a vortex-free trap region with subsequent smoothing; as an alternative approach, simultaneous combination of $C_{t=2}$ and $C_{s}$ terms is also successfully implemented to demand both accuracy and smoothness. Calculated outputs corresponding to these examples are illustrated in Fig.~\ref{fig:cterms} for the same second-order power-law pattern shown in Fig.~\ref{fig:vortices}.
	 
	For the sequential $C_{t}$ application shown in Fig.~\ref{fig:cterms}(a), we initially use $C_{t=4}$ to apply coarse corrections to the calculated intensity pattern, then follow this with more refined corrections using $C_{t=2}$.  The cost function on signal region pixels is given a weighting of 10 times that of noise region pixels during the $C_{t=4}$ stage; for $C_{t=2}$ application the signal region is weighted by a factor of $10^{12} (1+T)$ relative to the noise region with $T$ the target value of a given pixel, while a linear slope over 8 pixels smooths the weightings between these two regions and discourages noise accumulation near the target intensity distribution.  The calculated fractional RMS error is 1.4\% after 3000 iterations of $C_{t=4}$ application and 0.07\% following an additional 10000 iterations of $C_{t=2}$ smoothing, with 70 vortices remaining in the signal region. However, as they are confined to low-intensity regions these vortices do not degrade the trapping pattern. The efficiency of the algorithm in placing light within the trapping region is 47\% prior to smoothing and 45\% afterwards.  However, efficiency varies widely according to weighting choice: one method of improving efficiency is to demand an accuracy across the entire output plane comparable to that of the signal region, requiring more iterations to achieve the desired trapping region accuracy. 

In the combined $C_{t=2}$ and $C_{s}$ approach, the signal and noise regions are not given relative weightings for the $C_{t=2}$ term, but the $C_{s}$ term is weighted according to the target value for each pixel within the signal region and set to zero in the noise region.  Competition between the accuracy and smoothing terms reduce accuracy as compared to the pure discrepancy power method, with a fractional RMS error  of 0.43\% after 30000 iterations in the example shown in Fig.~\ref{fig:cterms}(b). However, appropriate balancing of terms results in effective vortex suppression and sufficient prediction accuracy. 130 vortices remain in the low-intensity region of the trapping pattern, which again is less relevant for atom trapping. Furthermore, regional weightings increase the efficiency to 64\%. We also find that this active smoothing method is particularly resilient to initialisation conditions, increasing the chance of success of a given iteration run. 

	\begin{figure}[!htb]
	\centering			
	{
		\includegraphics[scale=.35]{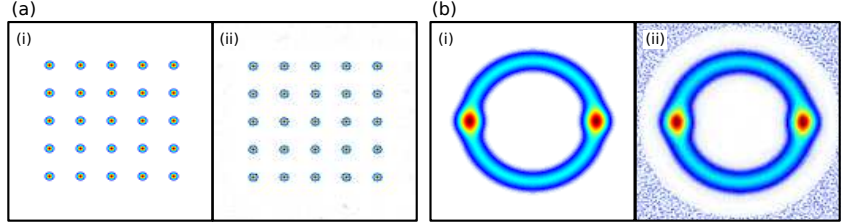}
		}
			\caption{Target (i) and calculated output intensity (ii) for the square lattice (a) and stirring ring pattern (b). Both are generated from a random initial phase and have RMS errors of 0.58\% and 3.0\% respectively.
                        }
			\label{fig:extrapot}
	\end{figure}

	With MRAF, the output quality depends critically on the initial phase guess and on the initialisation parameters, which have to be carefully chosen to suppress the formation of optical vortices during the calculation process~\cite{pasienski}. In contrast, with conjugate gradient minimisation, optical vortex suppression is achieved by a judicious cost function choice. Having determined these cost functions, the output quality is then largely insensitive to the initial phase guess, to the point that high accuracy can be achieved even with a random phase guess. For this reason, while conjugate gradient minimisation converges in more iterations than are required in MRAF, the two methods end up with a comparable computational efficiency, because with conjugate gradient minimisation it is not necessary to run the code for many different choices of initial phase patterns. Initialisation resilience would be of further benefit in dynamical sequences as it increases the chance of success of all frames from a single initialisation step.
	
	Displayed in Fig.~\ref{fig:extrapot} are examples illustrating the general applicability of the method to both continuous distributions and discrete arrays, showcasing the successful elimination of high-intensity borders next to the signal region. Both patterns are generated from a random initial guess; remarkably, for the ring pattern, we find that an initially random phase pattern results in quicker convergence to the final form than an apparently educated guess.  The lattice pattern is calculated using solely a $C_{t=2}$ term: smoothing is found to cause blurring of the pattern edges, and the $C_{t=2}$ term is sufficient to remove vortices from these smaller intensity features before they become established. The small size of the spots allows the vortices to escape more easily and they are therefore not frozen into the final pattern. The signal region is weighted by a factor of $10^{4}$ relative to the noise region, with a linearly sloped border of 4 pixels connecting these two regions sufficient to discourage noise accumulation near the signal region boundary. After 4000 iterations, the RMS signal region error is 0.58\%. The stirring ring pattern, so called because it can be used to induce superfluid rotation~\cite{Bruce_ring_2011}, has an RMS signal region error of 3.0\% after 2500 iterations. This example is calculated using a combination of $C_{t=2}$ and $C_{s}$ terms, with the signal region given an overall weighting of 10 relative to the noise region with a border of 8 pixels connecting these to prevent disruptive noise accumulation, and the smoothing term given a weighting of \textonehalf$(T_{max}-T_{nm})$, with $T_{max}$ the maximum target value and $T_{nm}$ the target values on individual pixels, in the signal region only. 
	
\section{Conclusion}

Conjugate gradient minimisation of an appropriate cost function has been verified as a viable alternative to the established methods of hologram generation.  By applying well-developed conjugate gradient approaches and optimised numerical libraries, the cost function minimisation approach allows careful guiding of the calculation process by choice of a sensible cost function with an analytical gradient. In particular, we show that tailoring the cost function beyond its simplest form is important and that the flexibility inherent in the cost function definition should be exploited to guide output plane features of interest.  We illustrate this by directly suppressing optical vortices in the calculated intensity profiles by associating a cost either with large deviations from the target intensity or with intensity fluctuations between neighbouring pixels.  Both methods successfully suppress vortices to optimise trapping potential accuracy, though active smoothing is more appropriate in potentials without sharp features.  This precision guiding and the ability to tailor the weightings assigned to each pixel has also allowed us to avoid the formation of the high-intensity signal region border characteristic of the MRAF approach, while retaining the flexibility characteristic of regional definitions and the ability to accurately reproduce a wide range of intensity patterns suitable for trapping ultracold atoms.  The method has also proven to be resilient to initialisation conditions, which may be of benefit in designing dynamic sequences of intensity patterns from a single initialisation step.
	
	Further improvements could be achieved by incorporating a measured laser beam profile into the calculation process to correct for beam imperfections, and considering Helmholtz propagation of light within the model~\cite{Gaunt_2012}. 

\nocite{*}

\end{document}